\documentclass[11pt]{article}

\usepackage{acl}
\usepackage{times}
\usepackage{latexsym}
\usepackage[T1]{fontenc}
\usepackage[utf8]{inputenc}
\usepackage{microtype}
\usepackage{inconsolata}
\usepackage{graphicx}
\usepackage{booktabs}
\usepackage{multirow}
\usepackage{array}
\usepackage{tabularx}
\usepackage{adjustbox}
\usepackage{makecell}
\usepackage{amsmath}
\usepackage{listings}   
\usepackage{xcolor}  
\usepackage{pifont}         
\usepackage[dvipsnames,table]{xcolor}
\usepackage{amssymb}
\usepackage{rotating}

\usepackage{xcolor}

\title{MMAG: A Multi-Control Mixed Audio Generation Benchmark}

\author{
  \textbf{Zihao Zheng}$^{1,2}$ \quad \textbf{Xuenan Xu}$^{2}$ \quad \textbf{Jiahao Mei}$^{1}$ \quad \textbf{Yixuan Li}$^{1,2}$ \\
  \textbf{Minghao Lv}$^{1}$ \quad \textbf{Wen Wu}$^{2}$ \quad \textbf{Chao Zhang}$^{2}$ \quad \textbf{Mengyue Wu}$^{1}$\thanks{Corresponding author.} \\[2pt]
  $^{1}$ MoE Key Lab of Artificial Intelligence, X-LANCE Lab, Shanghai Jiao Tong University, China \\
  $^{2}$ Shanghai AI Lab, China \\
  \texttt{rookie9@sjtu.edu.cn, mengyuewu@sjtu.edu.cn}
}
\begin{document}
\maketitle

\begin{abstract}

Recent audio generation systems have progressed from single-modality synthesis to generating complex acoustic scenes containing speech, music, and sound effects.
Therefore, evaluating these models requires assessing multiple interacting capabilities, including semantic fidelity, speaker consistency, and temporal control, yet existing benchmarks focus on isolated domains or coarse-grained descriptions.
To address this gap, we introduce the Multi-control Mixed Audio Generation (MMAG) benchmark.
MMAG contains approximately 4,000 manually verified audio clips with rich annotations covering speech content, speaker identity, music attributes, sound events, and temporal relationships, together with dedicated subsets for voice cloning and timestamp-conditioned generation.
We further propose a systematic evaluation protocol that measures acoustic fidelity, speech quality, semantic alignment, and temporal accuracy.
Benchmarking representative agentic orchestrators, unified audio-visual generation models, and native mixed-audio generators reveals substantial performance trade-offs across these capabilities, with no existing model performing consistently well.
Our results highlight the remaining challenges of controllable mixed audio generation and establish MMAG as a comprehensive benchmark for future research.\footnote{The project page is available: \url{https://hirookie9.github.io/MMAG-Page/}}

\end{abstract}

\section{Introduction}

Research in audio generation has historically progressed along three parallel tracks, targeting speech, music, and sound effects separately. 
Each track has developed its own specialized model architectures and datasets, with little cross-domain interaction~\citep{hung2026tangoflux, zhou2026indextts2, copet2023simple}.

With the development of generative models, several works have started to integrate different audio domains. These efforts fall into three categories: agentic orchestrators that compose scenes via expert models, unified audio-visual generators that produce synchronized video-audio, and native mixed-audio generators that unify speech, music, and effects in a single end-to-end system.
\begin{table*}[t]

\centering

\tiny
\caption{Comparison of MMAG with existing benchmarks. Domain columns indicate the audio categories covered. Condition columns indicate the types of annotations or control signals provided. \textcolor{green}{\ding{51}} indicates availability. \textcolor{red}{\ding{55}} indicates unavailability.}
\begin{tabular}{l r c c c c c c c c c c}   
\toprule
\multirow{2}{*}{Benchmark} & \multirow{2}{*}{Size} & \multicolumn{3}{c}{Domain} & \multicolumn{7}{c}{Condition} \\
\cmidrule(lr){3-5} \cmidrule(lr){6-12}
 & & Speech & Sound & Music & Transcript & Speaker Attr. & Sound Event & Music Info & Temporal Order & Voice Prompt & Timestamp \\
\midrule
LibriSpeech-PC &   1.2k & \textcolor{green}{\ding{51}} & \textcolor{red}{\ding{55}} & \textcolor{red}{\ding{55}} & \textcolor{green}{\ding{51}} & \textcolor{red}{\ding{55}} & \textcolor{red}{\ding{55}} & \textcolor{red}{\ding{55}} & \textcolor{red}{\ding{55}} & \textcolor{green}{\ding{51}} & \textcolor{red}{\ding{55}} \\
MusicBench & 0.4k & \textcolor{red}{\ding{55}} & \textcolor{red}{\ding{55}} & \textcolor{green}{\ding{51}} & \textcolor{red}{\ding{55}} & \textcolor{red}{\ding{55}} & \textcolor{red}{\ding{55}} & \textcolor{green}{\ding{51}} & \textcolor{red}{\ding{55}} & \textcolor{red}{\ding{55}} & \textcolor{red}{\ding{55}} \\
AudioCaps & 4.8k & \textcolor{green}{\ding{51}} & \textcolor{green}{\ding{51}} & \textcolor{red}{\ding{55}} & \textcolor{red}{\ding{55}} & \textcolor{red}{\ding{55}} & \textcolor{green}{\ding{51}} & \textcolor{red}{\ding{55}} & \textcolor{green}{\ding{51}} & \textcolor{red}{\ding{55}} & \textcolor{red}{\ding{55}} \\
MECAT & 20k & \textcolor{green}{\ding{51}} & \textcolor{green}{\ding{51}} & \textcolor{green}{\ding{51}} & \textcolor{green}{\ding{51}} & \textcolor{green}{\ding{51}} & \textcolor{green}{\ding{51}} & \textcolor{green}{\ding{51}} & \textcolor{red}{\ding{55}} & \textcolor{red}{\ding{55}} & \textcolor{red}{\ding{55}} \\
\midrule
\rowcolor{gray!10}
\textbf{MMAG (Ours)} & \textbf{4k} & \textcolor{green}{\ding{51}} & \textcolor{green}{\ding{51}} & \textcolor{green}{\ding{51}} & \textcolor{green}{\ding{51}} & \textcolor{green}{\ding{51}} & \textcolor{green}{\ding{51}} & \textcolor{green}{\ding{51}} & \textcolor{green}{\ding{51}} & \textbf{0.7k}\quad\textcolor{green}{\ding{51}} & \textbf{1.8k}\quad\textcolor{green}{\ding{51}} \\
\bottomrule
\end{tabular}
\label{tab:benchmark_comparison}
\end{table*}

The emergence of mixed audio generation models poses new challenges for existing benchmarks.
The first challenge lies in limited audio content and textual annotations.
Most existing benchmarks are designed for single-domain tasks: Text-to-Speech (TTS) test sets such as LibriSpeech-PC~\citep{meister2023librispeech} contain only clean speech, and Text-to-Music (TTM) benchmarks like Music-Bench~\citep{melechovsky2024mustango} consist solely of instrumental music, neither covering cross-domain content.
In contrast, Text-to-Audio (TTA) test sets such as AudioCaps~\citep{kim2019audiocaps} do include speech and sound elements but suffer from coarse captions that omit speech transcriptions and speaker details.
MECAT~\citep{niu2026mecat} produces richer captions for mixed scenes, yet it lacks temporal ordering of sound events for generation, which would require a more comprehensive annotation pipeline.

The second challenge is the absence of fine-grained control mechanisms in existing benchmarks, as users in practice demand not only high-quality audio but also precise controllability over specific attributes. 
One example is zero-shot voice cloning in TTS~\citep{zhou2026indextts2}, where the model mimics a speaker's timbre from a short voice prompt.
This necessitates that the benchmark provide suitable prompt samples. 
However, selecting such segments from real-world audio is challenging due to complex acoustic scenes.
Another example is temporal control in TTA, which requires the model to generate sound events at precisely specified timestamps~\citep{xie2025picoaudio}. 
This requires obtaining precise timestamps for each element during the annotation process.

To address these challenges, we introduce MMAG, a benchmark for Multi-control Mixed Audio Generation.
We construct three evaluation settings that differ in conditioning signals: (1) text-only control on the main set, (2) extra speaker identity control via a short voice prompt, and (3) temporal control via timestamp-detailed captions.
We select cross-domain audio samples from test sets of AudioCaps, VGGSound, and MECAT.
Extending the multi-expert annotation pipeline of MECAT with a timestamp-labeling branch, we construct detailed captions including speech transcriptions, speaker attributes (age, gender, dialect, and emotion), sound event descriptions, musical information (genre, instrument and mood) and temporal ordering.
After manual inspection, the main set, voice cloning subset, and timestamp subset comprise around 3,980, 690, and 1,800 audio-text pairs respectively, with the latter two providing reference voices and precise temporal boundaries accordingly.
Table~\ref{tab:benchmark_comparison} compares MMAG with existing benchmarks across coverage of domains and conditions.

Using MMAG, we systematically evaluate representative models from these emerging system families. 
To provide a holistic assessment of mixed audio generation, we establish an evaluation protocol across different dimensions: acoustic fidelity, speech quality, semantic consistency, and temporal controllability.
Our findings reveal that existing models struggle to achieve balanced performance across different metrics.
Moreover, introducing voice prompts leads to a noticeable performance drop, and most models exhibit little capability for fine-grained temporal control.

Our contributions are summarized as follows:
\begin{enumerate}
    \item We introduce \textbf{MMAG}, the first benchmark for \emph{compositional evaluation} of mixed audio generation. MMAG comprises a manually verified benchmark with rich annotations spanning speech, music, sound effects, and temporal relationships, together with dedicated subsets for prompt-based voice cloning and timestamp-conditioned generation.
    \item We establish a unified evaluation framework that jointly measures acoustic fidelity, speech quality, semantic consistency and temporal control, providing a holistic assessment of mixed audio generation beyond conventional single-metric evaluation.
    \item We benchmark representative agentic orchestrators, unified  audio-visual models, and native mixed-audio generators under this unified protocol, revealing that current systems exhibit substantial trade-offs across generation quality and controllability, with no single model consistently excelling across all evaluation dimensions.
\end{enumerate}

\section{Related Work}
\begin{figure*}[h!]
    \centering
    \includegraphics[width=\textwidth]{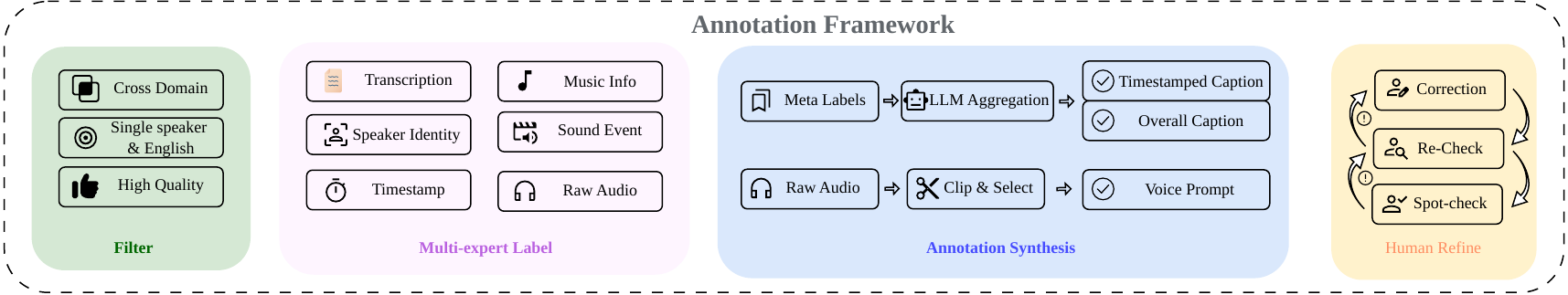}
    \caption{Overview of the MMAG annotation pipeline. Expert models extract fine-grained attributes from audio clips. An LLM aggregates these annotations into coherent overall and timestamped captions, while speech segments are further processed to construct voice prompts for voice cloning evaluation. Human inspection is performed to ensure annotation quality.}
    \label{fig:model}
\end{figure*}
\paragraph{Benchmarks for audio generation}

Existing benchmarks for audio generation include AudioCaps~\citep{kim2019audiocaps} and VGGSound~\citep{chen2020vggsound}, which provide large-scale real-world audio with coarse annotations.
and MECAT~\citep{niu2026mecat}, built from ACAV100M~\citep{lee2021acav100m}, which also serves as a valuable audio source and offers richer annotations through expert-model and LLM combination.
However, none of these resources are designed for compositional evaluation of mixed audio generation.
\paragraph{Evaluation of audio generation}
Existing metrics for audio generation have been developed along separate lines, each targeting a specific quality dimension. For distributional metrics, Fréchet Audio Distance (FAD)~\citep{kilgour2018fad}, Fréchet Distance (FD), Kullback-Leibler (KL) divergence, and Inception Score (IS)~\citep{salimans2016improved} are widely used to quantify how closely the generated audio matches the real distribution. Alongside these metrics, no-reference predictors such as Audiobox Aesthetics~\citep{tjandra2025audiobox} provide perceptual quality assessments across multiple axes. For speech metrics, Word Error Rate (WER), speaker similarity (SPK-SIM), and UTMOS\footnote{UTMOS-v2~\citep{baba2024t05}.} address transcription accuracy, identity preservation, and naturalness respectively. For semantic metrics, Contrastive Language-Audio Pretraining (CLAP) score~\citep{wu2023clap} provides global text-audio alignment, while AnyAudio-Judge~\citep{li2026anyaudio} achieves fine-grained semantic evaluation by decomposing captions into atomic rubrics.
While these metrics cover complementary aspects of quality, they remain scattered across isolated benchmarks, with no unified protocol for evaluating controllable mixed audio generation.
\paragraph{Audio generation systems}
Recent audio generation systems can be grouped into three paradigms.

\textit{Agentic orchestrators}, represented in our evaluation by AuDirector~\citep{ren2026audirector}, employ LLMs as central agents to invoke expert models including TangoFlux~\citep{hung2026tangoflux}, IndexTTS2~\citep{zhou2026indextts2}, and MusicGen~\citep{copet2023simple}. 
AuDirector further incorporates critic models such as CLAP and MiMo-Audio~\citep{xiaomi2025mimo} to iteratively evaluate and refine generated audio quality.
This training-free design enables flexible adaptation to diverse conditioning inputs, including voice references or timestamps.

\textit{Unified audio-visual generation models} accept diverse input modalities and synthesize synchronized audio-visual content.
Representative models in our evaluation include MOVA~\citep{yu2026mova}, Ovi~\citep{low2025ovi}, UniAVGen~\citep{zhang2025uniavgen}, LTX-2~\citep{hacohen2026ltx2}, and JavisDiT++~\citep{liu2026javisdit++}, which can synthesize synchronized video alongside general audio covering speech, music, and sound effects.

\textit{Native mixed-audio generators}, represented by Dasheng AudioGen~\citep{mei2026dasheng}, synthesize mixed audio scenes from text. 
Ming-Omni-TTS~\citep{inclusion2025mingomni}, a speech-centric system derived from TTS, extends its scope to joint audio-music generation.

These increasingly capable systems motivate a benchmark capable of evaluating multiple control dimensions simultaneously.

\section{Constructing MMAG}
MMAG is designed to enable \emph{compositional evaluation} of mixed audio generation, where speech, music, and sound effects must be generated coherently within a single acoustic scene. Rather than introducing another collection of audio-text pairs, our construction process is guided by three principles: (1) preserving realistic mixed-audio scenarios, (2) providing rich semantic and temporal annotations for fine-grained control, and (3) enabling multi-condition control through dedicated subsets.
Figure~\ref{fig:model} illustrates the overall construction process.
\begin{figure*}[h!]
    \centering
    \includegraphics[width=\textwidth]{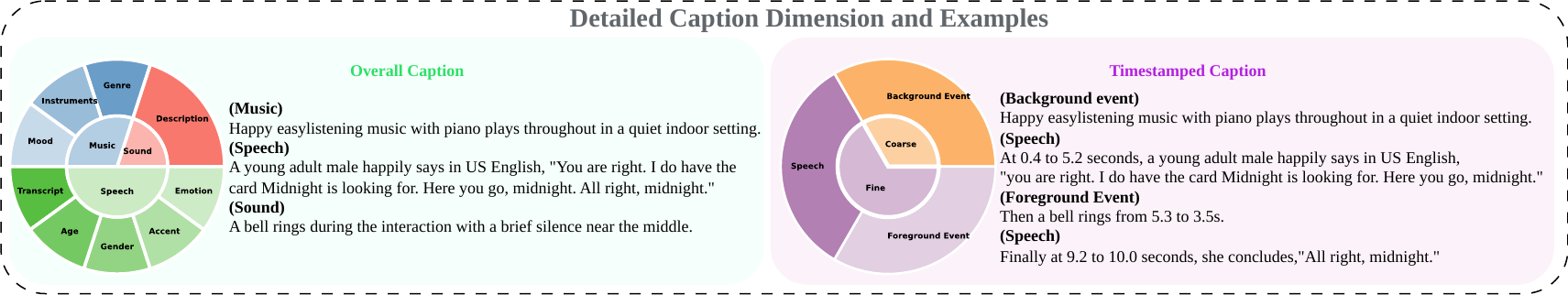}
    \caption{Illustrative example of two caption types for the same audio clip. Overall captions annotate speech, sound, and music, along with relative temporal order.
    Timestamped captions provide precise timestamps for foreground events and speech segments.}
    \label{fig:caption}
\end{figure*}

\subsection{Data Sources and Filtering}

Candidate clips are collected from test sets of AudioCaps, VGGSound, and MECAT, which provide complementary acoustic coverage for diverse speech, music, and environmental sound combinations while remaining comparable to widely used evaluation resources. 
We apply CED~\citep{dinkel2024ced} to detect the presence of these modalities and retain only clips that contain cross-domain content.
For clips containing speech, we further apply expert models to annotate speaker number and language, retaining only single-speaker English clips for reliable speech annotation.
We additionally filter out samples with poor recording quality or audible distortion using production quality (PQ) metric~\citep{tjandra2025audiobox} to ensure audio fidelity.

\subsection{Annotation Pipeline}

\paragraph{Fine-grained expert annotation.} 

For speech, expert models infer speaker gender, age, accent, emotion, and transcription.
For music, we annotate instrumentation, genre, and mood.
For sound events, we identify foreground and background sound classes along with ambient acoustic context.
In addition, we introduce a timestamping branch that provides precise temporal boundaries for all annotated elements across different domains.
\paragraph{Overall Caption.}

An LLM~\citep{deepseek2025r1} aggregates annotations from multiple expert models and synthesizes them into a single coherent caption.
The prompt instructs the model to faithfully retain expert-derived facts, reconcile contradictions, and avoid unsupported inferences.
The LLM further leverages timestamp information to describe the temporal progression of audio elements, using phrases such as “at the beginning,” “subsequently,” or “throughout.”
This temporal structuring is particularly important in TTA tasks, where constructing a coherent narrative of the acoustic scene is essential for generation quality.
\paragraph{Timestamped Caption.}
In addition to the overall description, we generate a separate timestamped caption for each clip, primarily intended for fine-grained temporal control evaluation.
This caption shares the same core semantic content as the overall one, but differs in precise timestamps for speech segments and foreground acoustic events.
At the event level, for sustained events such as background music, we retain only coarse temporal descriptions such as "throughout the audio."
At the clip level, we discard the timestamped caption entirely for any clip whose duration is almost entirely occupied by speech or background audio, since such samples offer limited scope for fine-grained temporal conditioning.

\paragraph{Voice Prompt.}

For audio samples from AudioCaps and VGGSound, we retrieve the original raw audio from YouTube.
We first identify the longest speech segment from the audio sample to serve as the speaker reference.
Separately, we apply voice activity detection (VAD)~\citep{gao2023funasr} to detect all speech segments in each raw audio. 
Then we crop or concatenate these segments to a suitable duration, generating a set of candidate prompts.
Finally, we use a speaker verification model~\citep{wang2023wespeaker} to compare each candidate against the reference, retaining the top-5 most similar for manual inspection.

\paragraph{Human quality control.}

To correct hallucinations introduced by model-based annotation, we perform manual inspection to ensure the quality.
This inspection focuses on two aspects: (1) the factual completeness and accuracy of the generated captions, and (2) the selection of final voice prompts from the candidate segments.
The manual inspection consists of three rounds: initial correction, re-check, and final spot-check.
The final spot-check pass rate reaches approximately $90\%$.
Given the inherent ambiguity in certain audio categories and temporal boundaries, the pass rate provides reasonable assurance of annotation quality.
\subsection{Benchmark Subsets}

The final benchmark comprises three subsets: a main set for comprehensive evaluation, and two additional subsets targeting specific capabilities.
An illustrative example of the two caption types is shown in Figure~\ref{fig:caption}.
\paragraph{Main set.}
The main set contains 3,974 annotated audio clips with manually verified overall captions. 
It serves as the primary evaluation set for acoustic fidelity, speech quality, and semantic consistency.

\paragraph{Voice Cloning subset.}
The voice cloning subset contains 691 clips, each paired with a voice prompt and a detailed overall caption.
Prompts are 2-5 seconds, matching the prompt length range in LibriSpeech-PC for zero-shot voice cloning.
This subset enables separate evaluation of speaker similarity in addition to the main-set metrics.

\paragraph{Timestamp subset.}
The timestamp subset contains approximately 1,828 clips selected from the main set, each annotated with fine-grained temporal boundaries for speech segments and foreground acoustic events.
This subset is specifically designed to evaluate temporal control in generation.

\section{Evaluation Protocol}

\subsection{Evaluated Models}

We evaluate the following models with their official checkpoints: unified audio-visual generation models (LTX-2.3, Ovi 1.1, MOVA-720p, JavisDiT++, UniAVGen), native mixed-audio generators (Dasheng-AudioGen, Ming-Omni-TTS), and the agentic orchestrator AuDirector.
Dasheng-AudioGen supports structured input for fine-grained control with improved transcript stability.
To this end, we evaluate both the unstructured and structured versions, denoted as Dasheng-AudioGen-Base and Dasheng-AudioGen-Fine.
For Ming-Omni-TTS, we evaluate two variants: 0.5B and 16.8B-A3B (denoted as 16.8B for brevity).

We evaluate all models on the main set using their respective supported input formats. 
For models that require a reference image (e.g., MOVA and UniAVGen), we provide a blank image to ensure uniform conditioning. 
JavisDiT++ does not accept transcript input, so we evaluate it without transcripts.
Ming-Omni-TTS requires structured text input, so we use an LLM to preprocess the caption into the required format.

On the voice cloning subset, we evaluate UniAVGen, Ming-Omni-TTS, and AuDirector.
We adapt AuDirector to use the provided prompt audio as input instead of retrieving a speaker embedding from its internal timbre library.

On the timestamp subset, we evaluate LTX-2, Ovi, MOVA, UniAVGen, Dasheng-AudioGen, and AuDirector, all of which support natural language input.
JavisDiT++ is excluded from this subset due to its lack of transcript support, while the subset contains extensive speech timestamps.
Ming-Omni-TTS is also omitted, as its structured input format is incompatible with timestamp control.
AuDirector is also extended with timestamp-based processing and synthesis capabilities for this subset.
\subsection{Automatic Metrics}

\paragraph{Acoustic Fidelity.}
We evaluate acoustic fidelity from two complementary perspectives: distributional similarity and perceptual quality.
For distributional similarity, we report Fréchet Distance (FD) in the PANNs CNN14 feature space~\citep{hershey2017cnn}, along with KL divergence and Inception Score (IS) computed using a pretrained audio classifier on AudioSet~\citep{gemmeke2017audioset} labels.
For completeness, we additionally provide FD scores computed with VGGish features in Appendix~\ref{app:results}, as the two FD variants exhibit similar trends.
For perceptual quality, we include Audiobox Aesthetics~\citep{tjandra2025audiobox}, a suite of no-reference predictors.
We report production quality (PQ) and production complexity (PC) as primary metrics, since audio quality and content complexity are core dimensions for mixed audio tasks.
Content enjoyment (CE) and content usefulness (CU) are deferred to the appendix~\ref{app:results}, as they are more subjective and application-dependent.

\paragraph{Speech Quality.}
For clips containing speech, WER is computed from Whisper-Large-v3~\citep{radford2023robust} transcriptions and capped at 1.0 for robustness.
For the voice cloning subset, we extract speaker embeddings using WeSpeaker-ResNet-221~\citep{wang2023wespeaker} and compute SPK-SIM.
$\text{UTMOS}_{\text{v2}}$~\citep{baba2024t05} estimates speech naturalness.
\begin{table*}[t]
\centering
\small
\begin{adjustbox}{max width=\textwidth}
\begin{tabular}{l*{10}{c}}   
\toprule
\multirow{2}{*}{Category} & \multirow{2}{*}{Model} & 
\multicolumn{5}{c}{\textbf{Acoustic}} & 
\multicolumn{2}{c}{\textbf{Speech}} & 
\multicolumn{2}{c}{\textbf{Semantic}} \\
\cmidrule(lr){3-7}
\cmidrule(lr){8-9}
\cmidrule(lr){10-11}
 &  & 
FD$\downarrow$ & KL$\downarrow$ & IS$\uparrow$ & PC$\uparrow$ & PQ$\uparrow$ & 
WER$\downarrow$ & UTMOS$_{\text{v2}}$$\uparrow$ & 
CLAP$\uparrow$ & AAJ$\uparrow$ \\
\midrule
\multirow{1}{*}{Agentic} 
& AuDirector & 5.19 & 1.60 & 2.62 & 3.69 & 6.71 & 0.08 & 2.68 & \underline{\textbf{0.34}} & 0.76 \\
\midrule
\multirow{5}{*}{Audio-Visual} 
& LTX-2 & 2.89 & 1.31 & 2.90 & 3.04 & 7.22 & 0.04 & 2.90 & \underline{\textbf{0.34}} & 0.76 \\
& Ovi & 3.91 & 1.50 & 2.06 & 2.60 & 6.05 & 0.13 & 2.25 & 0.30 & 0.74 \\
& MOVA & 4.38 & 1.28 & 2.59 & 2.98 & 6.74 & 0.15 & 2.69 & 0.31 & 0.78 \\
& JavisDiT++ & 5.58 & 1.47 & \underline{\textbf{3.49}} & \underline{\textbf{3.84}} & 5.98 & -- & 1.64 & 0.30 & 0.59 \\
& UniAVGen & 10.19 & 2.28 & 1.44 & 2.35 & 6.75 & 0.10 & 3.00 & 0.25 & 0.64 \\
\midrule
\multirow{4}{*}{Mixed-Audio} 
& Dasheng-AudioGen-Base & \underline{\textbf{1.78}} & \underline{\textbf{0.87}} & 3.17 & 3.55 & 6.67 & 0.11 & 2.77 & \underline{\textbf{0.34}} & \underline{\textbf{0.80}} \\
& Dasheng-AudioGen-Fine & 1.79 & 0.89 & 3.11 & 3.64 & 6.79 & 0.11 & 2.76 & 0.33 & \underline{\textbf{0.80}} \\
& Ming-Omni-TTS-0.5B & 8.72 & 2.80 & 1.56 & 1.60 & \underline{\textbf{7.35}} & \underline{\textbf{0.03}} & \underline{\textbf{3.43}} & 0.26 & 0.65 \\
& Ming-Omni-TTS-16.8B & 7.98 & 2.46 & 1.65 & 1.79 & 7.14 & \underline{\textbf{0.03}} & 3.33 & 0.27 & 0.67 \\
\bottomrule
\end{tabular}
\end{adjustbox}
\caption{Results on the main set.}
\label{tab:main}
\end{table*}
\paragraph{Semantic Consistency.}
CLAP score~\citep{wu2023clap} assesses semantic consistency at the global level by measuring the cosine similarity between the input caption and the generated audio in a contrastive embedding space.
To further assess semantic consistency at a fine-grained level, we employ AnyAudio-Judge~\citep{li2026anyaudio}, which first uses an LLM~\citep{yang2025qwen3} to decompose each caption into a set of atomic rubric items, each targeting a single dimension such as speaker age or music genre.
A multimodal LLM then scores the generated audio against each rubric independently.
For each rubric, the model produces a compliance probability derived from the softmax-normalized logits of the “yes” vs. “no” response.
The final sample score is obtained by averaging all rubric-level scores, denoted as AAJ.

 \paragraph{Temporal Control.}
For the timestamp subset, we compute segment-level F1 (Seg-F1) to evaluate temporal alignment.
The ground-truth timestamps for speech segments and foreground acoustic events are obtained from timestamped captions using an LLM.
For predictions, we apply WhisperX~\citep{bain2023whisperx} to extract speech timestamps and PE-A-Frame-Large~\citep{vyas2025pushing} to extract foreground event timestamps, both on the generated audio.
We then align predictions with ground truth and report $\text{Speech}_{\text{F1}}$ and $\text{Sound}_{\text{F1}}$ separately.

\begin{table*}[t]
\centering
\small
\begin{adjustbox}{max width=\textwidth}
\begin{tabular}{l*{11}{c}}   
\toprule
\multirow{2}{*}{Category} & \multirow{2}{*}{Model} & 
\multicolumn{5}{c}{\textbf{Acoustic}} & 
\multicolumn{3}{c}{\textbf{Speech}} & 
\multicolumn{2}{c}{\textbf{Semantic}} \\
\cmidrule(lr){3-7}
\cmidrule(lr){8-10}
\cmidrule(lr){11-12}
 & & 
FD$\downarrow$ & KL$\downarrow$ & IS$\uparrow$ & PC$\uparrow$ & PQ$\uparrow$ & 
WER$\downarrow$ & SPK-SIM$\uparrow$ & UTMOS$_{\text{v2}}$$\uparrow$ & 
CLAP$\uparrow$ & AAJ$\uparrow$ \\
\midrule
\multirow{1}{*}{Agentic} 
& AuDirector & \underline{\textbf{3.80}} & \underline{\textbf{2.11}} & \underline{\textbf{2.61}} & \underline{\textbf{3.54}} & 5.56 & 0.10 & 0.72 & 2.32 & \underline{\textbf{0.35}} & \underline{\textbf{0.75}} \\
\midrule
\multirow{1}{*}{Audio-Visual} 
& UniAVGen & 6.70 & 2.58 & 1.58 & 3.28 & 5.84 & 0.28 & 0.67 & 2.40 & 0.26 & 0.70 \\
\midrule
\multirow{2}{*}{Mixed-Audio} 
& Ming-Omni-TTS-0.5B & 8.72 & 3.32 & 1.26 & 1.81 & \underline{\textbf{6.54}} & 0.06 & 0.76 & \underline{\textbf{2.85}} & 0.26 & 0.66 \\
& Ming-Omni-TTS-16.8B & 7.39 & 2.99 & 1.38 & 2.12 & 6.43 & \underline{\textbf{0.04}} & \underline{\textbf{0.78}} & 2.79 & 0.27 & 0.68 \\
\bottomrule
\end{tabular}
\end{adjustbox}
\caption{Results on the voice cloning subset.}
\label{tab:sub}
\end{table*}

\begin{table*}[t]
\centering
\small
\begin{adjustbox}{max width=\textwidth}
\begin{tabular}{l*{12}{c}}   
\toprule
\multirow{2}{*}{Category} & \multirow{2}{*}{Model} & 
\multicolumn{5}{c}{\textbf{Acoustic}} & 
\multicolumn{2}{c}{\textbf{Speech}} & 
\multicolumn{2}{c}{\textbf{Semantic}} & 
\multicolumn{2}{c}{\textbf{Temporal}} \\
\cmidrule(lr){3-7}
\cmidrule(lr){8-9}
\cmidrule(lr){10-11}
\cmidrule(lr){12-13}
 & & 
FD$\downarrow$ & KL$\downarrow$ & IS$\uparrow$ & PC$\uparrow$ & PQ$\uparrow$ & 
WER$\downarrow$ & UTMOS$_{\text{v2}}$$\uparrow$ & 
CLAP$\uparrow$ & AAJ$\uparrow$ & 
Speech$_{\text{F1}}$$\uparrow$ & Sound$_{\text{F1}}$$\uparrow$ \\
\midrule
\multirow{1}{*}{Agentic} 
& AuDirector & 6.59 & 2.01 & 3.08 & \underline{\textbf{4.16}} & 6.19 & 0.15 & 2.31 & \underline{\textbf{0.35}} & 0.78 & \underline{\textbf{0.72}} & \underline{\textbf{0.57}} \\
\midrule
\multirow{4}{*}{Audio-Visual} 
& LTX-2 & 2.38 & 1.46 & 3.35 & 3.35 & \underline{\textbf{6.97}} & \underline{\textbf{0.05}} & 2.74 & 0.33 & 0.77 & 0.64 & 0.44 \\
& Ovi & 3.99 & 1.67 & 2.23 & 2.74 & 6.05 & 0.19 & 2.12 & 0.27 & 0.78 & \underline{\textbf{0.65}} & 0.39 \\
& MOVA & 4.36 & 1.82 & 2.91 & 2.85 & 6.69 & 0.14 & 2.47 & 0.29 & 0.77 & 0.63 & 0.42 \\
& UniAVGen & 11.18 & 2.52 & 1.39 & 2.66 & 6.56 & 0.15 & \underline{\textbf{2.90}} & 0.23 & 0.67 & 0.63 & 0.29 \\
\midrule
\multirow{2}{*}{Mixed-Audio} 
& Dasheng-AudioGen-Base & 1.96 & 1.10 & 3.67 & 3.47 & 3.48 & 0.43 & 2.68 & 0.31 & 0.78 & 0.60 & 0.41 \\
& Dasheng-AudioGen-Fine & \underline{\textbf{1.77}} & \underline{\textbf{1.09}} & \underline{\textbf{3.76}} & 3.49 & 6.64 & 0.13 & 2.65 & 0.31 & \underline{\textbf{0.80}} & 0.62 & 0.42 \\
\bottomrule
\end{tabular}
\end{adjustbox}
\caption{Results on the timestamp subset.}
\label{tab:timestamp}
\end{table*}

\section{Results and Analysis}

We report results on the main set, voice cloning subset, and timestamp subset in Tables~\ref{tab:main}, \ref{tab:sub}, and \ref{tab:timestamp}. For all tables, arrows ($\uparrow/\downarrow$) indicate the direction of better performance, and best results per column are bolded and underlined.

\subsection{Main Set Results}

Table~\ref{tab:main} compares all evaluated models on the main set.
The most salient observation is that no model dominates across different metrics simultaneously.

Dasheng-AudioGen obtains the strongest distributional audio scores.
Dasheng-AudioGen-Base achieves the best FD (1.78) and KL (0.87) among the evaluated systems, while Dasheng-AudioGen-Fine delivers nearly identical FD/KL performance with slightly higher perceptual quality scores.
This result confirms Dasheng-AudioGen's superior performance in terms of acoustic fidelity among the evaluated models.
Its high CLAP and AAJ scores further demonstrate strong semantic alignment and fine-grained consistency with the input captions.
However, its speech metrics are not dominant: both variants obtain WER = 0.11, lagging behind the leading models on speech intelligibility.

Ming-Omni-TTS shows the opposite trend.
Both the 0.5B and 16.8B variants achieve the best WER (0.03) on the main set, and the 0.5B model obtains the highest UTMOS$_{\text{v2}}$ and PQ among all models.
This confirms the advantage of speech-specialized generation for intelligibility and perceived speech naturalness.
However, Ming-Omni-TTS shows a systematic disadvantage on distribution indicators, PC and semantic metrics, ranking well below the average. 
As a speech-centric system, it handles only a limited set of label-formatted sound events. 
However, the diversity and richness of non-speech content in MMAG exceed its capability.
Consequently, it frequently omits music and sound effects, which degrades acoustic and semantic scores while inflating speech-related metrics.

Among audio-visual generation models, LTX-2 delivers the most balanced performance on the main set.
It achieves strong FD, competitive KL and IS, the best CLAP score, and nearly the lowest WER. 
Its PQ is also high, second only to Ming-Omni-TTS-0.5B.
This suggests that unified audio-video models can transfer useful scene-generation abilities to audio-only evaluation even when visual input is held constant.
MOVA and Ovi show moderate performance across most metrics, while UniAVGen underperforms acoustic fidelity despite achieving reasonable UTMOS$_{\text{v2}}$, largely because it omits most music and sound events in its generations.
JavisDiT++ achieves the highest IS and PC, implying diverse and acoustically complex outputs, but its speech is largely unintelligible, resulting in limited performance on speech quality metrics.
At fine grained semantic consistency level, AAJ scores align with this trend.
MOVA, LTX-2, and Ovi achieve similar scores (0.74-0.78). 
UniAVGen (0.64) and JavisDiT++ (0.59) score considerably lower, reflecting their respective weaknesses in sound/music generation and speech intelligibility.

The agentic orchestrator AuDirector achieves a CLAP score tied for the best, along with competitive PC and AAJ score, yet it lags behind the leading systems on other metrics.
Although AuDirector can generate nearly all constituent audio elements, the final synthesis suffers from poor volume equalization and abrupt transitions between acoustic events.
This results in perceptually disjointed outputs that differ from real acoustic distributions and consequently degrade both distribution and speech metrics.

\subsection{Voice Cloning Subset Results}

Table~\ref{tab:sub} evaluates the voice cloning subset, where introducing voice prompts degrades several metrics across models, albeit to varying degrees.
For UniAVGen, WER rises from 0.10 on the main set to 0.28 on the voice cloning subset, while UTMOS$_{\text{v2}}$ drops from 3.00 to 2.40 and PQ from 6.75 to 5.84.
Ming-Omni-TTS shows a similar trend: although it remains the strongest system for both WER and SPK-SIM, its UTMOS$_{\text{v2}}$ and PQ are noticeably lower than those on the main set. AuDirector follows the same pattern, with both WER and PQ degrading.

These results indicate that speaker voice conditioning introduces a distinct control challenge: the model must not only generate the requested acoustic content, but also preserve speaker identity from a short prompt and maintain intelligibility.
SPK-SIM confirms that models extract useful identity information from the prompt: Ming-Omni-TTS-16.8B achieves the highest SPK-SIM of 0.78, followed by Ming-Omni-TTS-0.5B at 0.76 and AuDirector at 0.72.
However, these high similarity scores do not guarantee uniformly better audio quality.
The model with the best SPK-SIM does not achieve the best performance on acoustic and semantic indicators. This highlights a trade-off between speaker identity preservation and full-scene generation quality.

Overall, AuDirector achieves strong performance on semantic metrics and PC score, highlighting the advantage in content planning and semantic preservation.
Its strong SPK-SIM also demonstrates good generalizability to voice cloning.
However, its most notable weakness lies in audio quality, as indicated by the lowest UTMOS$_{\text{v2}}$ and PQ among all models.
UniAVGen exhibits the most pronounced degradation on this subset, suggesting that balancing and integrating multimodal information remains a challenge for unified audio-visual generation models.
Ming-Omni-TTS continues to show strong speech capability alongside poor sound event generation, mirroring its main-set performance.

\subsection{Timestamp Subset Results}

Table~\ref{tab:timestamp} evaluates fine-grained temporal control.
Current models struggle to follow timestamped instructions for both speech and acoustic events.
Even the best-performing system, AuDirector, reaches Speech$_{\text{F1}}$ = 0.72 and Sound$_{\text{F1}}$ = 0.57, leaving a substantial gap to reliable temporal-controllable synthesis.
Unified Audio-visual models and Dasheng-AudioGen show near-random timestamp controllability, while the other metrics trends mirror the main set.

Among existing approaches, AuDirector achieves the highest temporal scores, benefiting from its specific adaptation for timestamp input and time-aware synthesis. 
However, this temporal controllability comes at a substantial fidelity cost, as AuDirector has greater degradation on FD and UTMOS$_{\text{v2}}$ relative to its main-set performance than other models.
In addition, the event detectors used for evaluation often fail to locate precise temporal boundaries in AuDirector's outputs, lowering both Speech$_{\text{F1}}$ and Sound$_{\text{F1}}$.
This indicates that more sophisticated synthesis strategies are needed to reduce the fidelity cost of temporal conditioning.

Beyond AuDirector, most other models also exhibit performance degradation across acoustics and speech metrics on the timestamp subset.
The decrease in indicators can be attributed to two factors. 
First, the timestamp input may fall outside the training distribution of existing models.
Second, the timestamp subset itself contains more complex and acoustically diverse audio content.

Beyond model performance, the automatic metrics exhibit a limitation on this subset.
Specifically, AAJ scores exhibit minimal dispersion across models.
The rubrics include timestamp-aware items, but the judge model exhibits limited discriminative ability on them, leading to undifferentiated scores.
This limitation points toward the need for more temporally aware judge models in future work.
\subsection{Error Patterns}

We observe that partial models exhibit caption confusion: they tend to produce spoken content drawn from the caption but outside the target transcript, even with correct input formatting.
To quantify this phenomenon, we design a diagnostic metric \textbf{Proportion} in Appendix~\ref{app:results}, defined as the contribution of such extraneous content to the final WER.
A higher Proportion indicates more severe leakage, independent of the absolute WER level.
\begin{table}[h]
\centering
\small
\caption{Proportion ($\%$) on the main/timestamp set. DASG: Dasheng-AudioGen; Ming: Ming-Omni-TTS.}
\begin{tabular}{l c c}
\toprule
\textbf{Model} & \textbf{Pro. (Main)} & \textbf{Pro. (Timestamp)} \\
\midrule
AuDirector   & \textbf{\underline{1.88}} & \textbf{\underline{5.43}} \\
\midrule
LTX-2           & 5.12  & 7.41 \\
Ovi             & 13.92 & 19.70 \\
MOVA            & 4.59  & 7.00 \\
UniAVGen        & 3.75  & 6.63 \\
\midrule
DASG-Base       & 18.21 & 56.48 \\
DASG-Fine       & 21.88 & 36.04 \\
Ming-0.5B       & 3.78 & -- \\
Ming-16.8B      & 2.80 & -- \\
\bottomrule
\end{tabular}
\label{tab:proportion}
\end{table}

The results in Table~\ref{tab:proportion} show that this issue is particularly pronounced in Ovi and Dasheng-AudioGen. On the main set, roughly 10$\%$-20$\%$ of word errors are attributable to caption confusion. On the timestamp subset, this problem is further amplified. Dasheng-AudioGen-Base exhibits a Proportion of 56.48$\%$, indicating a strong correlation between increased WER and caption confusion. After adopting structured input, Dasheng-AudioGen-Fine shows substantial reductions in both WER and Proportion, though its Proportion remains higher than that of other models.

Compared to the main set, all models show increased Proportion on the timestamp subset.
This suggests that the introduction of timestamps raises the difficulty of recognizing and generating accurate transcripts, highlighting the need for more diverse training data. 
Dasheng-AudioGen-Fine benefits from LLM-based input processing, which significantly reduces the confusion, and AuDirector also attains a strong Proportion score via LLM-driven input processing and expert model invocation.
Leveraging LLMs to normalize raw inputs into a form better suited for models represents an effective strategy to improve model robustness.

\section{Conclusion}

To address the lack of comprehensive evaluation for mixed audio generation, we propose MMAG.
It comprises a main set of approximately 4,000 clips, a timestamp subset of roughly 1,800 clips, and a voice cloning subset of roughly 690 clips. 
MMAG is curated through a multi-stage pipeline that integrates expert models, LLM-based caption synthesis, and human quality control.
We have established a comprehensive evaluation framework and systematically benchmarked existing models.
Our findings reveal that no existing model achieves balanced performance across acoustic fidelity, speech quality, semantic consistency, and temporal control.
Instead, we observe systematic trade-offs between generation quality and controllability, calling for more robust and controllable generation.

\section*{Limitations}
Despite its comprehensive scope, MMAG has several limitations.
The 10-second clip duration, single-speaker English restriction, and imbalanced sound/music categories limit generalizability and may bias fine-grained evaluation on underrepresented classes.
In addition, even with a multi-expert pipeline and manual verification, sound events, music genres, and precise timestamps remain inherently subjective, introducing unavoidable uncertainty into the ground truth.
For evaluation, we incorporate LLM-based semantic metrics, but current multimodal LLMs still show limited reliability on fine-grained audio tasks.
This is expected to improve as foundation models continue to advance.
\section*{Ethical Considerations}

We hired external annotators for manual quality control of generated captions and voice prompt selection. All annotators were compensated at a rate above local minimum wage. Annotators received detailed instructions and were informed of their right to withdraw at any time. No personally identifiable information was collected from annotators, and the annotation task did not involve sensitive content. All audio data used in this benchmark is sourced from existing publicly available datasets (AudioCaps, VGGSound, MECAT) and is used for research purposes in accordance with their original licenses. We do not foresee direct negative societal impacts from this work; the benchmark is intended to facilitate transparent and reproducible evaluation of audio generation models.

\bibliography{references}
\clearpage
\appendix

\section{Implementation Details}
\label{app:model}
To obtain high-quality annotations for each audio clip, we employ specialized expert models for each attribute dimension. Table~\ref{tab:attributes_models} summarizes the complete set of models used in our annotation pipeline.

For speaker number estimation, we adopt a conservative strategy by leveraging two complementary models: \textsc{Pyannote-SD 3.1} and \textsc{Audioflamingo3}. We take the minimum of the two predictions as the final speaker number, as preliminary experiments revealed that both models tend to overestimate the number of speakers.

\begin{table*}[htbp]
\centering
\caption{Attribute and Corresponding Models for Each Category}
\label{tab:attributes_models}
\begin{tabular}{>{\raggedright\arraybackslash}p{1.8cm} 
                >{\raggedright\arraybackslash}p{2.8cm} 
                >{\raggedright\arraybackslash}p{8cm}}
\toprule
\textbf{Category} & \textbf{Attribute} & \textbf{Model} \\
\midrule
\multirow{8}{*}{speech} 
& \multirow{2}{*}{speaker number} & Pyannote-SD 3.1~\citep{plaquet2023powerset} \\
& & Audioflamingo3~\citep{ghosh2026audio} \\
& language & Speechbrain-ECAPA~\citep{ravanelli2024open} \\
& age\&gender & Audeering-AGR~\citep{burkhardt2023speech} \\
& accent & CommonAccent~\citep{zuluagagomez23_interspeech} \\
& emotion & Emotion2Vec~\citep{ma2024emotion2vec} \\
& transcript & Whisper-Large-V3~\citep{radford2023robust} \\
& timestamp & WhisperX~\citep{bain2023whisperx} \\
\midrule
\multirow{3}{*}{sound} 
& \multirow{2}{*}{event} & CED~\citep{dinkel2024ced}\\
& & Gemini-2.5-pro~\citep{comanici2025gemini} \\
& timestamp & PE-A-Frame-Large~\citep{vyas2025pushing} \\
\midrule
\multirow{2}{*}{music} 
& instrument\&genre & Musical Descriptor~\citep{li2024mert} \\
& mood & Music2Emo~\citep{kang2025towards} \\
\bottomrule
\end{tabular}
\end{table*}
\section{LLM Prompt for Caption}
\label{app:prompt}

To generate the descriptive captions for our benchmark, we design the following instruction prompt for the LLM. The prompt enforces strict constraints on factual consistency, listening-only perspective, and structured hierarchical output. To improve information completeness, the model is required to first generate a content-focused caption for each category. The complete prompt is presented below.

\begin{lstlisting}
You are an expert AI audio analyst. Your task is to synthesize captions using the given "audio information" in JSONL format.

========================================
I. GENERAL RULES
========================================
a) Carefully identify conflicting information between fields and avoid mentioning conflicting aspects in the final caption. Focus only on consistent and unopposed information. Do not invent details not present in the data.

b) Do not use parentheses to provide detailed explanation in any output (e.g., avoid "Middle age (30-40 years old)").

c) All answers must be created from the perspective of someone who ONLY LISTENED to the audio, without any technical/model references or quantitative metrics (e.g. MOS).

========================================
II. CAPTION DEVELOPMENT FRAMEWORK
========================================

A. Content-Focused Caption
----------------------------------------
Generate a detailed description that focuses on individual content dimensions. Not all categories below may be present; if a field is missing, leave it blank or mark as unavailable. Do not hallucinate.

1) Speech:
   - Incorporate transcription content and speaker attributes (age, gender, language, accent, emotion, etc.).
   - Infer speaker identity only when sufficient evidence exists.
   - The complete transcription must be included verbatim, enclosed in quotation marks, without omission or summarization.
   - Example: "A middle-aged male captain aged 40-50 calmly stated in standard English, 'The plane is about to land.'"

2) Music:
   - Synthesize music descriptions based on the provided information (instrument, genre, moods).
   - Key, instrument, and genre are given as label-probability pairs. Integrate all relevant information to make informed judgments. 
   - The existing music labels have limited credibility and should only serve as auxiliary references for resolving ambiguous cases.
   - Example: "Slow-tempo electronic music with dark melodic elements and synthesized bass."

3) Sound:
   - Generate a complete description of sound events based on the "event_segments" field.
   - Sound events exclude content from major categories such as speech or music (e.g., "man talking", "guitar").
   - Include coarse-grained temporal information, such as event sequences and approximate ranges, but avoid precise timestamps.
   - Example: "After a barking dog, the sound of a baby crying could be heard, with wind mixed in the background."

4) Environment:
   - Infer environmental context from the existing speech, sound, and music information, and provide a refined environment caption when evidence is sufficient.
   - Example: "Indoor/outdoor hybrid environment with significant engine interference."

B. Systematic Caption
----------------------------------------
Generate holistic captions that integrate all content dimensions.

1) Overall Caption (several sentences):
   - Include as much information as possible, covering speech, music, sound, environment, and other relevant dimensions.
   - If transcription exists, include the complete transcription verbatim, enclosed in quotation marks. The speech content may be summarized at the beginning, but the complete transcription must appear subsequently.
   - Example: "A 20-30 young male speaker delivers an emotionally charged English monologue expressing pride, 'I love my country deeply'. Accompanied by dark-toned electronic music with steady percussion. Vehicle engine noise persists throughout the recording."

2) Timestamped Caption (several sentences):
   - Include as much information as possible, covering speech, music, sound, environment, and other relevant dimensions.
   - If transcription exists, include the complete transcription verbatim, enclosed in quotation marks.
   - Provide specific timestamps for each sentence and foreground event by integrating transcription-segments and event_segments data. Identify all foreground events, merging those with similar categories or highly overlapping time intervals.
   - For background sound events, do not provide precise timestamps; instead, describe their occurrence using coarse-grained terms such as "start", "end", "middle", or "entire process".
   - Example: "In 0.2-1.5s, a 20-30 young female speaker delivers an emotionally charged English monologue expressing pride, 'I love my country deeply'. Then applause erupted between 2.1-5.9 seconds. In 7.5-9.4s, the speaker said, 'my speech is over, thank you all.' There is dark electronic music and stable percussion in the background."
\end{lstlisting}

\section{Dataset Details}
\label{app:datasets}
\begin{figure*}[htbp]
    \centering
    \includegraphics[width=\textwidth]{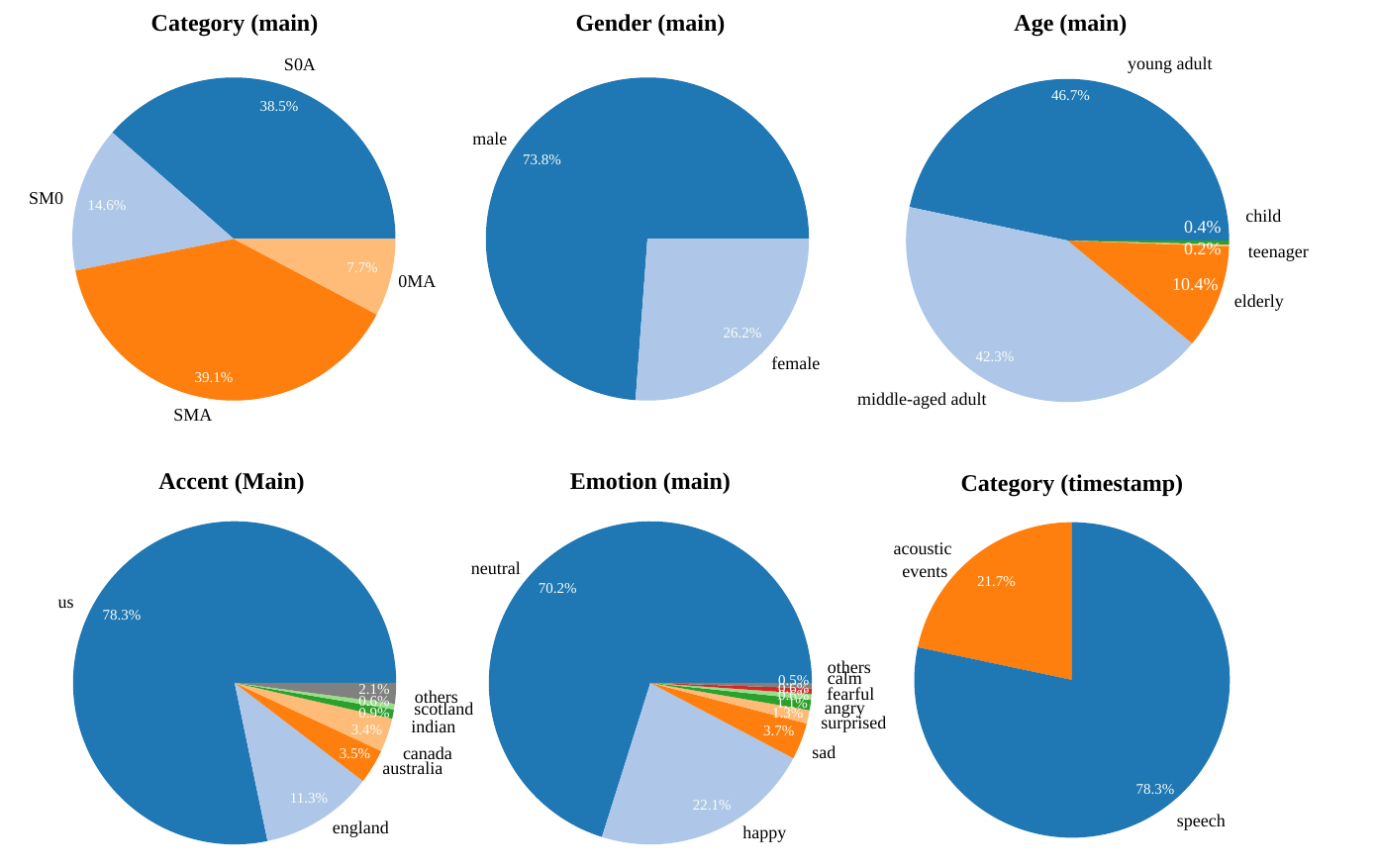}
    \caption{Label distribution of MMAG}
    \label{fig:label_distribution}
\end{figure*}

We perform a comprehensive statistical analysis of the benchmark to characterize its composition and coverage. Figure~\ref{fig:label_distribution} presents the label distributions on the main set across multiple dimensions, including gender, age group, emotion type, and accent category. 
For the timestamp subset, we report the proportion of samples with precise temporal annotations, covering both speech segments and foreground acoustic events.

\begin{figure}[htbp]
    \centering
    \includegraphics[width=0.45\textwidth]{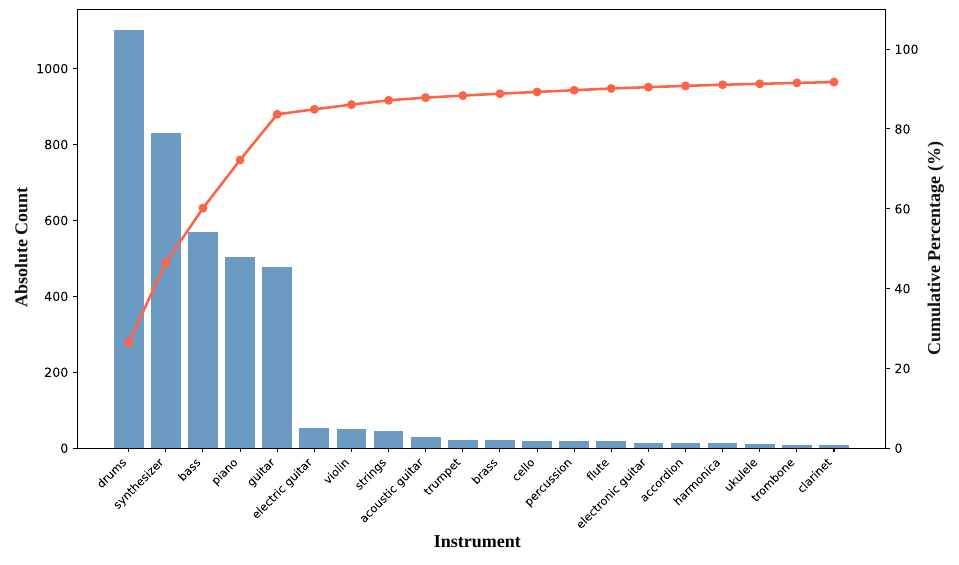}
    \caption{Instrument distribution of MMAG}
    \label{fig:instrument_distribution}
\end{figure}

To examine musical diversity, we extract instrument information from the generated captions using an LLM. Figure~\ref{fig:instrument_distribution} presents the per-clip instrument count distribution, revealing the varying instrumental complexity across samples.

For sound events, we extract event types from the captions via the same LLM, yielding roughly 320 distinct categories after excluding speech- and music-related events. 
In comparison, after removing 'speech,' 'music,' and their descendants from the AudioSet ontology, roughly 360 categories remain.
The substantial overlap between MMAG and AudioSet demonstrates that our benchmark achieves broad coverage of diverse sound events. 
\begin{figure}[htbp]
    \centering
    \includegraphics[width=0.45\textwidth]{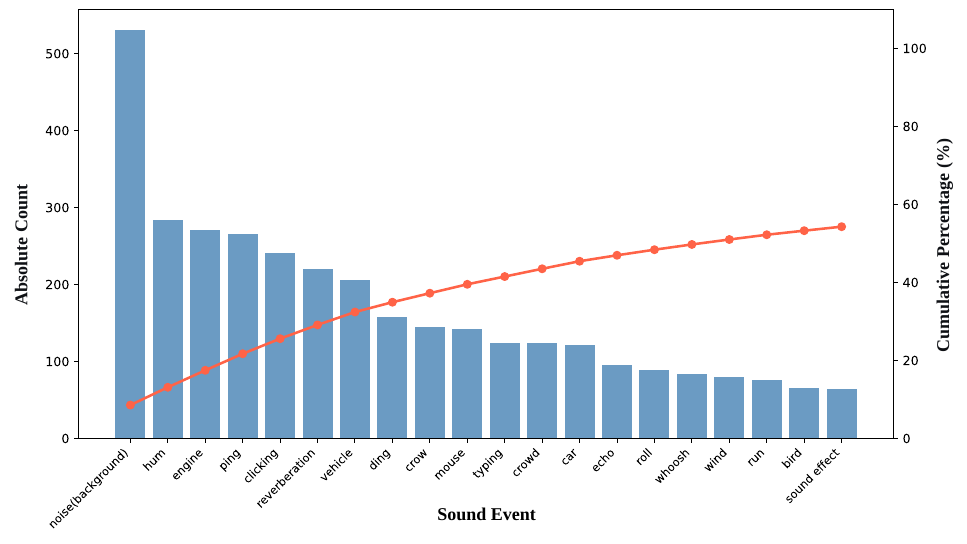}
    \caption{Sound event distribution of MMAG}
    \label{fig:event_distribution}
\end{figure}
\begin{figure}[htbp]
    \centering
    \includegraphics[width=0.45\textwidth]{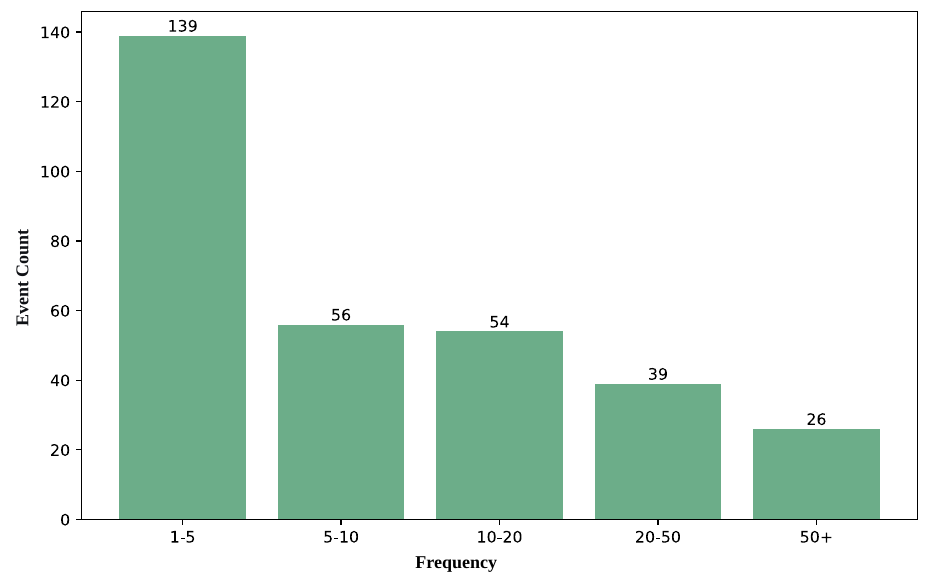}
    \caption{Frequency distribution of sound events}
    \label{fig:event_frequency}
\end{figure}
However, as shown in the Figure~\ref{fig:event_distribution} and Figure~\ref{fig:event_frequency}, the distribution of sound events in MMAG is not perfectly balanced.
This skew largely reflects the natural prevalence of certain event types in real-world recordings, inherited from the underlying data sources.
While this imbalance presents challenges for evaluating performance on rare categories, it also more faithfully mirrors the long-tail distribution characteristic of real acoustic environments.

For the voice cloning subset, we compute the SPK-SIM between each voice prompt and its corresponding target audio sample, achieving an average score of approximately 0.82. By comparison, the reference score computed on the LibriSpeech-pc dataset is approximately 0.86.

The slightly lower performance on our subset can be attributed to the presence of background acoustic content in some of the extracted voice prompts, which inevitably interferes with speaker embedding extraction and reduces the similarity score. 
However, such background-accompanied prompts are inherently reasonable in the context of mixed audio generation, where the model is expected to extract and preserve the target speaker's timbre not only from perfectly clean speech, but also from realistic, non-isolated audio segments.
From the SPK-SIM evaluation perspective, WeSpeaker remains generally robust to background speech and environmental noise. While not perfect in complex acoustic scenes, the metric still provides a meaningful measure of speaker similarity.

\section{Additional Results}
\begin{table*}[h!]
\centering
\footnotesize
\setlength{\tabcolsep}{4pt} 
\caption{Comparison across three benchmark subsets. Best results within each subset are highlighted in \textbf{bold} and underlined.}
\label{tab:exp_results_all}
\begin{tabular}{@{} l l c c c c c c c @{}}
\toprule
\multirow{2}{*}{Category} & \multirow{2}{*}{Model} & \multicolumn{3}{c}{Aux. Metric} & \multicolumn{4}{c}{WER Any.} \\
\cmidrule(lr){3-5} \cmidrule(lr){6-9}
                         &                       & FD$_{\text{CNN14}}$$\downarrow$ & CE$\uparrow$ & CU$\uparrow$ & Hit Rate (\%)$\downarrow$ & WER$_{\text{caps}}$ (\%)$\downarrow$ & WER (\%)$\downarrow$ & Proportion (\%)$\downarrow$ \\
\midrule
\multicolumn{9}{c}{\textbf{Main Set}} \\
\midrule
\multirow{1}{*}{Agentic}
& AuDirector             & 27.12                & 5.25   & 6.18   & 2.40    & 0.15    & 8.00    & \textbf{\underline{1.88}} \\
\midrule
\multirow{5}{*}{Audio-Visual}
& LTX-2                  & 17.09                & 5.88   & 6.70   & 1.64    & 0.19    & 3.71    & 5.12 \\
& Ovi                    & 22.47                & 5.31   & 6.66   & 12.35   & 1.87    & 13.43   & 13.92 \\
& MOVA                   & 18.56                & 5.62   & 6.21   & 8.69    & 0.69    & 15.03   & 4.59 \\
& JavisDiT++               & 21.05                & \textbf{\underline{4.11}} & \textbf{\underline{4.85}} & --       & --       & --       & --       \\
& UniAVGen               & 46.80                & 5.34   & 5.96   & 5.03    & 0.36    & 9.61    & 3.75 \\
\midrule
\multirow{4}{*}{Mixed-Audio}
& Dasheng-AudioGen-Base  & \textbf{\underline{10.46}} & 5.47   & 6.05   & 8.28    & 2.03    & 11.15   & 18.21 \\
& Dasheng-AudioGen-Fine  & 11.27                & 5.55   & 6.20   & 8.98    & 2.44    & 11.15   & 21.88 \\
& Ming-Omni-TTS-0.5B     & 82.31                & 5.90   & 6.97   & 1.39    & 0.11    & 2.94    & 3.74 \\
& Ming-Omni-TTS-16.8B    & 64.81                & 5.83   & 6.85   & \textbf{\underline{1.12}}    & \textbf{\underline{0.07}} & \textbf{\underline{2.50}} & 2.80 \\
\midrule \midrule
\multicolumn{9}{c}{\textbf{Voice Cloning Set}} \\
\midrule
\multirow{1}{*}{Agentic}
& AuDirector             & \textbf{\underline{40.33}} & \textbf{\underline{4.48}} & 5.10   & 2.19    & 0.23    & 10.00   & \textbf{\underline{2.30}} \\
\midrule
\multirow{1}{*}{Audio-Visual}
& UniAVGen               & 50.08                & 4.60   & \textbf{\underline{4.95}} & 12.41   & 1.59    & 27.50   & 5.78 \\
\midrule
\multirow{2}{*}{Mixed-Audio}
& Ming-Omni-TTS-0.5B     & 117.27               & 5.45   & 6.26   & 1.90    & \textbf{\underline{0.16}}    & 6.08    & 2.63 \\
& Ming-Omni-TTS-16.8B    & 92.57                & 5.41   & 6.10   & \textbf{\underline{1.01}}    & 0.26    & \textbf{\underline{4.48}}    & 5.80 \\
\midrule \midrule
\multicolumn{9}{c}{\textbf{Timestamp Set}} \\
\midrule
\multirow{1}{*}{Agentic}
& AuDirector             & 34.71                & \textbf{\underline{4.55}} & 5.57   & 7.57    & 0.82    & 15.10   & \textbf{\underline{5.43}} \\
\midrule
\multirow{4}{*}{Audio-Visual}
& LTX-2                  & 19.10                & 5.56   & 6.41   & \textbf{\underline{2.97}}    & \textbf{\underline{0.40}}    & \textbf{\underline{5.40}}    & 7.41 \\
& Ovi                    & 24.63                & 5.08   & \textbf{\underline{5.53}} & 20.82   & 3.68    & 18.68   & 19.70 \\
& MOVA                   & 27.52                & 5.31   & 6.14   & 7.08    & 1.01    & 14.42   & 7.00 \\
& UniAVGen               & 50.56                & 5.14   & 5.63   & 9.15    & 1.00    & 15.09   & 6.63 \\
\midrule
\multirow{2}{*}{Mixed-Audio}
& Dasheng-AudioGen-Base  & \textbf{\underline{13.43}} & 5.22   & 5.88   & 70.26   & 24.02   & 42.53   & 56.48 \\
& Dasheng-AudioGen-Fine  & 13.83                & 5.23   & 6.03   & 15.71   & 4.62    & 12.82   & 36.04 \\
\bottomrule
\end{tabular}
\end{table*}
\label{app:results}
In addition to the primary metrics, we report supplementary results for three auxiliary metrics (FD$_{\text{CNN14}}$, CE, and CU) in the supplementary tables (Tables~\ref{tab:exp_results_all}) for each of the three benchmark subsets.

FD$_{\text{CNN14}}$ computes the Fréchet distance between feature embeddings extracted from a pre-trained CNN14 model, measuring the distributional similarity between generated and reference audio. CE and CU are aesthetic predictors from the Audiobox suite. These metrics either partially overlap with the primary metrics in the main paper or are not central to the core evaluation focus. We therefore include their full results in the appendix for completeness.

\textbf{WER Any. metrics:} To assess the quality of predicted transcriptions, we design a specialized analysis targeting a specific error type: cases where the model generates content from the non-transcription parts of the reference caption (e.g., environmental descriptions, music annotations, or sound event descriptions) rather than the spoken transcript only. We define three metrics for this error type:

\begin{itemize}
\item \textbf{Hit Rate (\%)}: The percentage of samples containing at least one such error.
\item \textbf{$\text{WER}_{\text{caps}}$}: The average WER computed exclusively on samples that contain the aforementioned error type.
\item \textbf{Proportion}: The ratio of $\text{WER}_{\text{caps}}$ to the overall WER across all samples, i.e., $\text{Proportion} = \text{WER}_{\text{caps}} / \text{WER}$. A higher Proportion indicates that this error type contributes more substantially to the overall transcription error, signaling a more severe issue.
\end{itemize}

These metrics enable diagnostic analysis of whether a model tends to mistakenly incorporate non-speech textual content into its transcription output, providing a direct measure of its ability to generate only the spoken content as instructed.
\end{document}